\documentclass[12pt]{article}
\usepackage{graphicx}
\usepackage{epsfig}
\begin{document}
\def\1{\'{\i}}

\centerline{\Large\bf Stationary Oscillations in a Damped Wave Equation}
\centerline{\Large\bf from Isospectral Bessel Functions}
\vspace*{1cm}

\centerline{ N. Barbosa-Cendejas and M.A. Reyes}
\vspace*{2mm}

\centerline{\small\it Instituto de F\1sica, Universidad de
Guanajuato,\\ Apdo. Postal E143, 37150 Le\'on,Gto., M\'exico}

\begin{center}
\small
\begin{minipage}{14cm}

Using the isospectral partners of the Bessel functions
derived by Reyes {\it et al.} \cite{marco}, we find, on one hand, that these
functions show non-typical supersymmetric (SUSY) behavior and, on the
other, that the isospectral partner of the classical wave equation is
equivalent to that of a damped system whose oscillations do not
vanish in time, but show a non-harmonic shape.

\vspace*{3mm}


\vspace*{3mm}

Usando las compa\~{n}eras isoespectrales de las funciones de Bessel
obtenidas por Reyes {\it et al.} \cite{marco}, encontramos, por un
lado, que estas funciones muestran un comportamiento
at\1pico de SUSY, mientras que, por otro lado, la compa\~{n}era isoespectral
de la ecuaci\'on de onda cl\'asica es equivalente a la de un sistema
amortiguado cuyas oscilaciones no desvanecen con el tiempo, sino que
obtienen una forma que no es arm\'onica.

\vspace*{3mm}

{\it Descriptores}: Supersimetr\1a; ecuaci\'on de Bessel.

\vspace*{3mm}


\end{minipage}
\end{center}

\normalsize

\section{Introduction}

In quantum mechanics (QM), the number of problems which can be exactly solved
is very limited, and one can only hope to get the approximated solution  using
a variety of methods, or turn to a modified problem which can be described in
terms of the known exact problems.  In fact, one of the main virtues of
supersymmetric (SUSY) quantum mechanics is that there one can find an infinite
number of one-parameter problems  which posses the same spectra as the known
exact ones.  If SUSY cannot be applied, one can still find isospectral
solutions by using the classical factorization method, or the interwinning
method, to look for new equations with the old spectra.

In classical mechanics, the only problem which can be directly compared to
quantum mechanics is that described by the classical wave equation
\begin{equation}
\left( \nabla^2 - \frac{1}{v^2} \right) \psi(x,y;t) = 0 \, ,
\end{equation}
but it has been stated before that there is no SUSY partner of this equation
\cite{bagchi}. In fact, a factorization of the Bessel equation {\it a la} Infeld
and Hull cannot be find \cite{infyhul}.  However, the Bessel equation,
\begin{equation}
\frac{d^2 J_{n}(r)}{dr^2} + \frac{1}{r} \frac{dJ_{n}(r)}{dr} +
\left( 1- \frac{n^2}{r^2} \right) J_{n}(r) =0 \, ,
\label{besseleq}
\end{equation}
with $n\geq 0$, which arises from the wave equation after separation of
variables, still possess a factorization in terms of raising and lowering
operators defined by the equations \cite{gradsh}
\begin{equation}
A^+_n J_{n}(r) = \left( \frac{d~}{dr} - \frac{n}{r} \right ) J_{n}(r) =
- J_{n+1}(r) \, ,
\end{equation}
\begin{equation}
A^-_{n+1} J_{n+1}(r) = \left( \frac{d~}{dr} + \frac{n+1}{r} \right ) J_{n+1}(r)
= J_{n}(r) \, ,
\end{equation}
respectively.

Following the work of Mielnik \cite{bogd}, who finds a family of potentials
which posses the same spectrum as that of the one dimensional harmonic
oscillator, and the work of Pi\~na \cite{pina} on the factorization of some
special functions found in mathematical physics, Reyes {\it et al.}
\cite{marco} are able to derive second order differential equations which are
`isospectral' to the equations described in the Sturm-Liouville theory.  Note
that since the `spectral' parameter $n$ appears in $A^+_n$ and $A^-_n$,
contrary to  SUSYQM \cite{bogd}, it shows up in the partner equation in a more
complicated fashion than the original equation.

As one can see, obtaining the isospectral partner of the Bessel equation is
the first step toward an isospectral classical wave equation, this being the
main purpose of this letter.  But, before proceeding in this direction, we first
show that the partner functions of the Bessel functions show a unique and
unusual SUSY behavior.  Then, we show that the isospectral classical wave
equation that comes from the isospectral partner of the Bessel equation
resembles the problem of damped waves, which nevertheless do not vanish in
time, but show non-harmonic shapes due to this damping term.


\section{Isospectral Bessel Equation}

In ref.\cite{marco}, the isospectral partner of the Bessel equation was found as
%
\begin{equation}
 \frac{d^2{\widetilde J}_{n+1}}{dr^2}+
\frac{1}{r} \frac{d{\widetilde J}_{n+1}}{dr}+
\left(1-\frac{(n+1)^{2}}{r^{2}}\right) {\widetilde
J}_{n+1}=
-\frac{4n}{r^2}
\frac{\left[(2n+1)\gamma r^{2n}+1\right]}{\left(\gamma r^{2n}+1\right)^{2}}
~{\widetilde J}_{n+1},
\label{Besselisos}
\end{equation}
where, as is usual in SUSYQM, $n\geq 0$ and the lowest lying eigenvalue is lost.
Here, $\gamma$ is the parameter of the partner functions, with
$0\leq\gamma<\infty$.

The term on the right of eq.(\ref{Besselisos}) corresponds to the extra
potential function
term in a typical SUSYQM problem, with the difference that in this case, the
integral defining this term is exactly solvable.  This also happens in the
partner Bessel functions, which were found to be
\begin{equation}
{\widetilde J}_{n+1}(r;\gamma) = -J_{n+1}(r)+
\frac{2n}{r}\left(\gamma r^{2n}+1\right)^{-1} J_{n}(r) \, .
\end{equation}

Notice that the fact that these functions are explicitely found allows one to
take a closer look to their properties, and by doing so we are here able to
show that there exist a non-typical SUSY behavior for the isospectral Bessel
functions, in the following sense.

\begin{figure}[h!]
\begin{picture}(500,180)(-10,0)
 \thicklines


 \put(90,10){\dashbox(20,18){$\tilde \psi_0$}}
 \put(150,10){\framebox(20,18){$\psi_0$}}

 \put(145,28){\vector(-3,2){30}}
 \put(90,50){\framebox(20,18){$\tilde \psi_1$}}
 \put(150,50){\framebox(20,18){$\psi_1$}}

 \put(145,68){\vector(-3,2){30}}
 \put(90,90){\framebox(20,18){$\tilde \psi_2$}}
 \put(150,90){\framebox(20,18){$\psi_2$}}

 \put(145,108){\vector(-3,2){30}}
 \put(90,130){\framebox(20,18){$\tilde \psi_3$}}
 \put(150,130){\framebox(20,18){$\psi_3$}}

 \put(100,165){$\vdots$}
 \put(160,165){$\vdots$}


 \put(260,10){\dashbox(20,18){$\tilde J_0$}}
 \put(320,10){\framebox(20,18){$J_0$}}
 \put(315,28){\vector(-1,2){30}}

 \put(260,50){\framebox(20,18){$\tilde J_1$}}
 \put(320,50){\framebox(20,18){$J_1$}}
 \put(315,58){\vector(-1,0){30}}
 \put(315,68){\vector(-1,2){30}}

 \put(260,90){\framebox(20,18){$\tilde J_2$}}
 \put(320,90){\framebox(20,18){$J_2$}}
 \put(315,98){\vector(-1,0){30}}

 \put(260,130){\framebox(20,18){$\tilde J_3$}}
 \put(320,130){\framebox(20,18){$J_3$}}
 \put(315,138){\vector(-1,0){30}}

 \put(270,165){$\vdots$}
 \put(330,165){$\vdots$}

\end{picture}
\caption{Typical SUSY isospectral function generating scheme (left),
and the way Bessel isospectral partners are generated (right), in terms of
the $n\!-\!1$ and $n\!+\!1$ orders of the original functions.}
\label{fig1}
\end{figure}
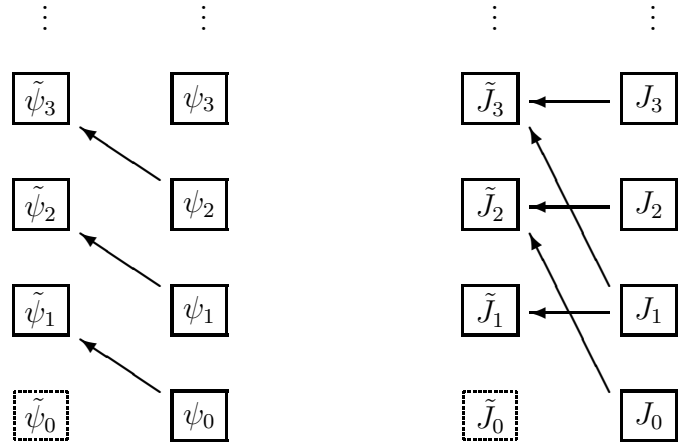

In ref.\cite{marco} it was thought that the partner functions where not
regular at $r\!=\!0$.  This is not the case, though.  One can see that
${\widetilde J}_1=-J_1$, and that for $n>0$ we can use one of
the recursion relations among Bessel functions, to say
\begin{equation}
\frac{2n}{r} J_n(r) = J_{n+1}(r)+J_{n-1}(r) \, ,
\end{equation}
to find that the partner Bessel functions are regular at $r\!=\!0$, since then
\begin{equation}
{\widetilde J}_{n+1}(r;\gamma) =
\frac{-\gamma r^{2n}}{\gamma r^{2n}+1}
J_{n+1}(r)+\frac{1}{\gamma r^{2n}+1} J_{n-1}(r) \, .
\label{partnbess}
\end{equation}
Now, this is a very unique feature of the Bessel partner functions.  In SUSYQM
the partner function of order $n+1$, $\tilde\psi_{n+1}(x;\gamma)$, relate to
the original function of preceding order $\psi_{n}(x)$ through a differential
operator.  Here, the partner
function of order $n+1$ relate to two Bessel functions, of orders
$n-1$ and $n+1$ (see fig.(\ref{fig1}).) Moreover, one can see that the limits
of the parameter $\gamma$ give
\begin{equation}
{\widetilde J}_{n+1}(r;\gamma\!=\!0) = J_{n-1}(r) \, ,
\label{jn-i}
\end{equation}
while
\begin{equation}
{\widetilde J}_{n+1}(r;\gamma\!=\!\infty) = -J_{n+1}(r) \, .
\end{equation}
Therefore, the partner Bessel function $\widetilde J_{n+1}(r;\gamma)$,
transforms from $J_{n-1}(r)$ to $-J_{n+1}(r)$ as $\gamma$ goes from 0 to
$\infty$.  This is its most unusual characteristic, to be related to a Bessel
function of two orders less than its own.

\section{Stationary oscillations in a damped wave equation}

In order to connect the partner Bessel funcions to the wave equation, we begin
writing eq.(\ref{Besselisos}) for $kr$ instead of $r$,
\begin{equation}
 r^2\frac{d^2{\widetilde J}_{n+1}}{dr^2}+
 r \frac{d{\widetilde J}_{n+1}}{dr}+
\left(k^2r^2-(n+1)^{2}\right) {\widetilde
J}_{n+1}= g_{n+1}(kr;\gamma){\widetilde J}_{n+1} \, ,
\label{polarbessel}
\end{equation}
where
\begin{equation}
g_{n+1}(u;\gamma)=
-\frac{4n \left[(2n+1)\gamma u^{2n}+1\right]}{\left(\gamma u^{2n}+1\right)^{2}}
 \, .
\end{equation}

Now, we multiply by a function of the polar angle $\theta$, $H(\theta)$, and
assume that this function satisfies the equation
\begin{equation}
\frac{d^2H}{d\theta^2} + (n+1)^2 H = 0 \, ,
\end{equation}
in order to write the wave equation
\begin{equation}
 \left(
 \nabla^2 - \frac{1}{v^2}
 \right) {\widetilde \psi}_{n+1}(r,\theta;t;\gamma) =
 \frac{g_{n+1}(kr;\gamma)}{r^2} \,
 {\widetilde \psi}_{n+1}(r,\theta;t;\gamma) \, ,
\label{U}
\end{equation}
where
${\widetilde \psi}_{n+1}(r,\theta;t;\gamma) =
{\widetilde J}_{n+1}(kr;\gamma) H(\theta)  e^{i\omega t}$.

This is the wave equation derived from the isospectral Bessel functions of
ref.\cite{marco}, possessing a damping term ${g_{n+1}(kr;\gamma)}/{r^2}$
which, though singular at $r=0$, allows stationary solutions whose radial part
is the partner Bessel function (\ref{partnbess}).
It is the kind of equation that may help on our comprehension of
neuronal activity, where such equations appear \cite{rosudwe}.

\begin{figure}[h!]
\hfill \epsfig{figure=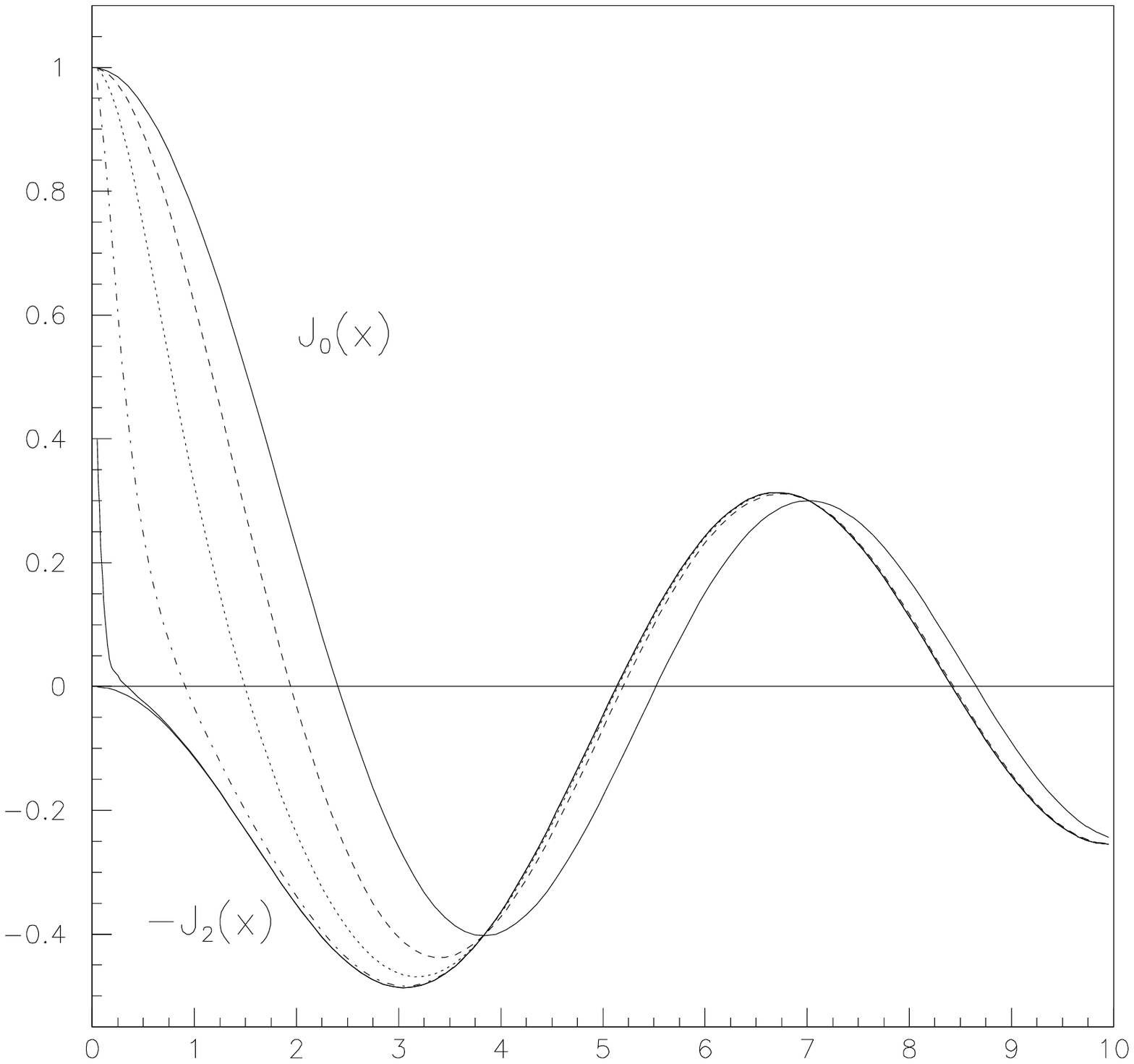, height=5.5cm}
\hfill
\epsfig{figure=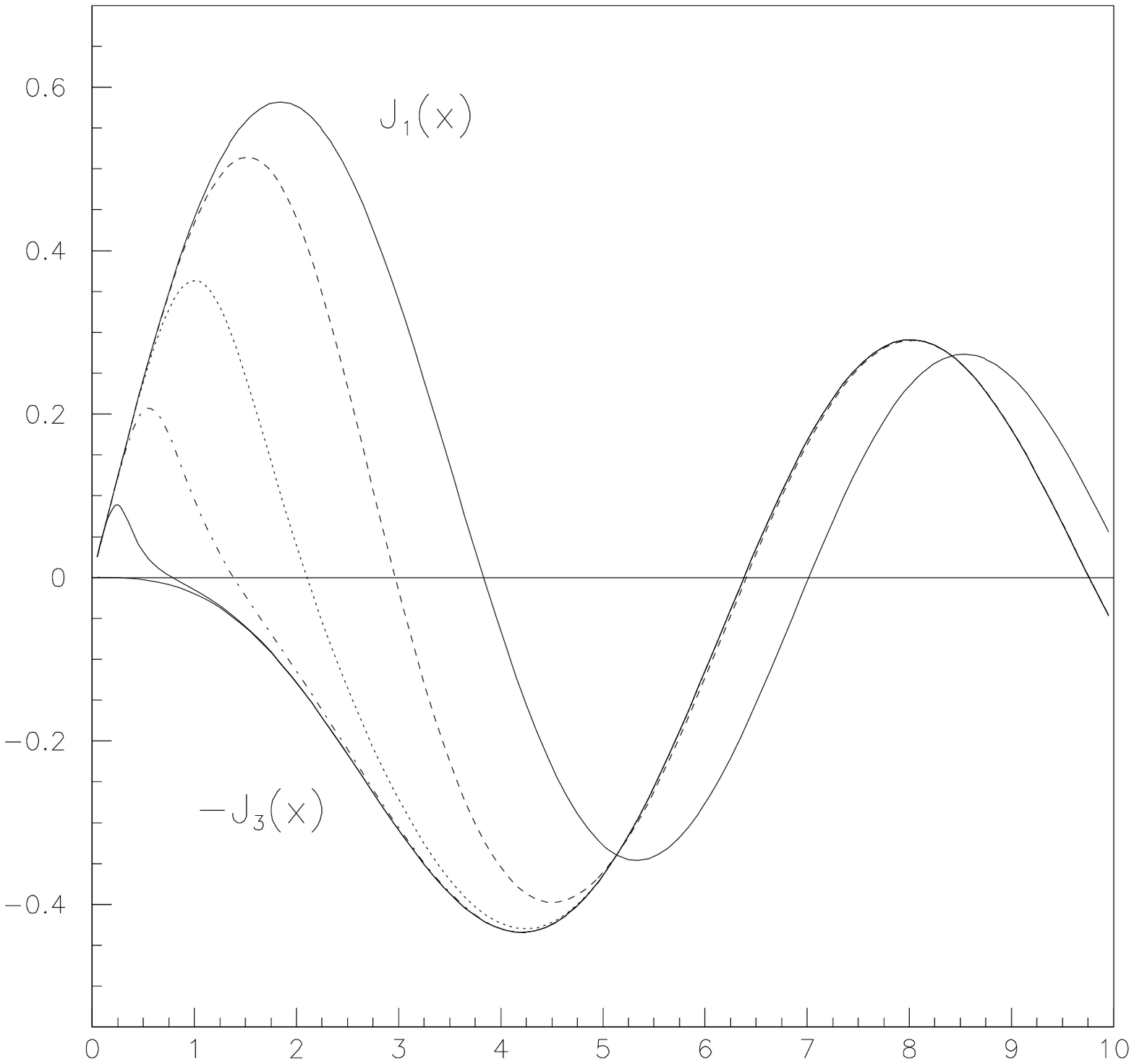, height=5.5cm}
\hfill
\caption{Partner Bessel functions: ${\widetilde J}_{2}$ (left),
evolving from $J_0$, $\gamma=0$, to $-J_2$, $\gamma=\infty$, and
${\widetilde J}_{3}$, evolving from $J_1$ to $-J_3$ (right). Note
how the damping term affects the typical harmonic shapes of the
Bessel functions, except when $\gamma=0$ and $\gamma=\infty$.}
\label{fig2}
\end{figure}

One may ask then, how does this damping term reflects in the stationary waves?
The answer comes from our discussion above, about the way $\widetilde J_{n+1}$
is determined by $J_{n-1}$ and $J_{n+1}$.  Notice that for $\gamma=0$ the
damping term reduces to $4n/r^2$ and is absorbed into the ordinary Bessel
equation (\ref{besseleq}), and that for $\gamma=\infty$ the damping term
reduces to zero. For other values of $\gamma$, the damping term modifies the
Bessel functions, making them lose their typical harmonic shapes, as
can be seen in fig.(2), where drastic changes in the harmonic shapes are seen as
$\gamma$ increases towards infinity, specially for $r$ close to zero.


\section{Conclusion}

In this letter we have shown that it is possible to find an isospectral
partner of the classical wave equation.  By using the isospectral partners of
the Bessel functions from ref.\cite{marco}, we have shown that the arising wave
equation contains a damping term, whose action does not destroy the stationary
waves, but makes them acquire non-harmonic shapes.  Since the Bessel equation
can not be supersymmetrized \cite{bagchi}, this may be the only way one can find
an isospectral partner of the wave equation.


\section{Acknowledgements}

We ackwoledge support form CONACYT of Mexico, through a scholarship for NBC, and
through grant No. SEP-2003-C0245364.


\end{document}